\documentstyle[preprint,aps]{revtex}
\begin{document}
\draft
\title{
Level Correlations for Metal-Insulator Transition}
\author{Pragya Shukla$^{*}$} 
\address{Department of Physics,   
Indian Institute of Technology, Kharagpur, India.}

\maketitle
\begin{abstract}

	We study the level-statistics of a disordered system undergoing 
 the Anderson type  metal-insulator transition. The disordered Hamiltonian 
 is a sparse random matrix in the site representation and the statistics 
 is obtained by taking an ensemble of such matrices. It is shown that the 
 transition of levels due to change of various parameters e.g. disorder,
 system size, hopping rate can be mapped to the perturbation driven 
 evolution of the eigenvalues  of an ensemble subjected to Wigner-Dyson 
type perturbation with an initial state given by a Poisson ensemble; the 
mapping is then used to obtain desired level-correlations.

\end{abstract}
.
 	The degree of disorder and dimensionality of a disordered system 
plays a significant role in the metal-insulator (MI) transition 
\cite{and}. The level 
statistics during the transition is governed by a parameter which is a 
function of both, disorder as well as dimensionality \cite{ss}. 
The purpose of this 
paper is to identify the parameter and study 
the effect of its variation on the spectral fluctuations. 

       The statistical behaviour of the levels of a disordered Hamiltonian 
depends on their number $r$  
within the energy ranges of interest and the dimensionless conductance $g$. 
The parameter $g$ is the ratio of two energy scales, namely, Thouless 
energy $E_c$ and mean level spacing $\Delta$. The energy scale 
$E_c=\bar h D/L^2$, with $D$ as diffusion coefficient and $L$ system size, 
corresponds to ergodic time-scales (ergodic classical limit) 
above which the wavefunction has fully 
diffused in the entire volume $L_d$ of the system and therefore does not 
feel the space dimensionality $d$. The eigenfunctions associated with levels
within  energy ranges smaller than $E_c$ are extended structureless 
objects leaving no signature on the level correlations. The number $r < g$,  
therefore, results in a 
universality of level correlations which can be well-modeled by the 
invariant ensembles of random matrix theory (RMT) \cite{fh}. For $r > g$, the 
time scales associated with the dynamics are such that the wavefunction 
is not yet fully diffused in the whole volume. The eigenfunctions in the 
corresponding energy-range are localized, either partially or fully 
depending on the dimensionality, and their spatial correlations can 
significantly affect the level correlations.

	The metallic phase of a disordered system corresponds to 
$g \rightarrow \infty$ in the thermodynamic limit $L\rightarrow 
\infty$.  
 The states, therefore, in any large but finite interval 
are always extended with a Wigner-Dyson (WD) level-statistics, characterstic 
of strong level-repulsion \cite{fh}.  
In the insulator phase 
$g \rightarrow 0$ in the thermodynamic limit, implying no overlap 
between eigenfunctions.  
 The corresponding level statistics can be modeled by the statistics 
of independent random numbers, that is, Poisson statistics.  
 A change in the degree of disorder results in the variation of $g$ 
(for fixed $r$) which can induce a transformation of extended states into 
localized states thereby causing a transition of spectral fluctuations.  
However the MI transition is very sensitive to the dimensions of the system. 
 For $d=1,2$ and, in thermodynamic 
limit, the system always remains an 
insulator even for very weak disorder.  
For $d > 2$, the increase of disorder results in decreasing 
$g$, leading to an extended to localized state transition at a 
critical value of disorder. 
In the region near the critical point of the transition, known as Anderson 
transition, 
the eigenstates are neither 
localized nor extended but form  self-similar multifractal structures 
 with strong fluctuations at all length scales\cite{ckl}. This 
results in a new type of universal level-statistics, known as 
critical spectral statistics (CSS),  which is neither Poissonian 
nor Wigner-Dyson but is a hybrid of both \cite{ss,klaa}.  
 For finite system size also,  
 a crossover 
from Poisson to WD statistics can be seen among levels as a function of 
the parameter $\zeta/L$ with $\zeta$ as the correlation length.
However the detailed analytical information about various level correlations 
in the critical regime for inifinte systems as well as for the 
intermediate stages, during 
transition in finite systems, is still missing. Our study attempts to 
bridge this gap.

	The level statistics for various transition stages  
is significantly affected by the  nature of the localization of the 
eigenfunctions 
and can suitably be modeled by an ensemble containing 
eigenvalue-eigenfunction 
correlations (therefore basis-dependent). 
Recently a new technique has been developed to deal with such cases
  where the eigenvalue distribution for an ensemble,  non-invariant under 
a change of basis, is mapped to the one for an ensemble invariant 
under change of basis and subjected to a random perturbation \cite{ps}. 
The mapping 
is achieved by identifying a basis-dependent parameter in the former 
to the perturbation (e.g symmetry-breaking) parameter in the latter. 
The distribution for 
the invariant case can 
subsequently be mapped to the non-stationary states of the particles 
governed by Calogero-Sutherland-Moser (CSM) Hamiltonian \cite{fh}. The known   
 particle correlations can therefore be used to determine level-correlations 
of the non-invariant ensembles.
 The present study uses the technique to analyze the 
spectral properties of the levels undergoing MI transition by taking  
 an ensemble of Anderson Hamiltonains in 
the site representation. As expected, the basis dependent 
parameter, governing the transition in this case, is a function of 
disorder, dimensionality and hopping rate and is related to $g$.

	The Anderson model for a disordered system is described by  
 a d-dimensional disordered lattice, of size $L$, with a 
  Hamiltonian $H= \sum_n \epsilon_n a_n^+ a_n + 
\sum_{n\not=m} t (a_n^+ a_m +a_n a_m^+)$ in tight-binding 
approximation \cite{and}. 
Here $a_n^+$ and $a_n$ are the 
creation and annihilation operators of an electron at a site $n$ in a 
lattice and $m$ refers to the nearest neighbors 
of the site $n$. 
The site energies $\epsilon_n$, measured in units of the 
overlap integral between adjacent sites correspond to the random potential. 
  The hopping is generally 
assumed to connect only the $z$ nearest-neighbors with amplitude $t$ so that 
the electron kinetic energy spread or bandwidth is $zt$.   
In the 
configration space representation  the Hamiltonian $H$  
turns out to be a sparse matrix of size $N=L^d$ with diagonal 
matrix elements as the site-energies $\epsilon_i$.   
 The level-statistics can 
therefore be studied by analyzing the properties of an ensemble of (i) sparse 
real symmetric matrics in presence of a time-reversal symmetry  and 
(ii) sparse complex Hermitian matrice in absence of a time-reversal 
symmetry.

We first consider 
the case of MI transition brought about by decreasing the diagonal 
disorder. In this case, site-energies $\epsilon_i$ 
are taken to be independent random variables with 
probability-density $p(\epsilon_i)$. In the original Anderson model \cite{and} 
 $p(\epsilon)$ was taken to be a constant $W^{-1}$ between 
$-W/2$ to $W/2$. Various physical arguments and approximations used in this 
case led to 
conclusion that all the states are localized for 
$W > 4 Kt {\rm ln}({W\over 2 t})$ with $K$ as a function of $z$ and $d$.   
However, as  well-known now,  MI transition does not depend 
on the nature of $p(\epsilon)$ and latter can also be chosen as Gaussian; 
 the type of $p(\epsilon)$ affects only the critical point of 
the transition \cite{and}.  
The ensemble measure $\rho(H)$, for any intermediate state of MI 
transition brought about  by diagonal disorder, can therefore be described 
by $ \rho (H,y,b)\propto 
{\rm exp}({- \sum_{s=1}^2 \sum_{k} \alpha_{kk;s}
 (H_{kk;s}- b_{kk;s})^2 }) \prod_{k\le l; s=1,2}\delta(H_{kl;s} - b_{kl;s})$   
 with subscript $"s"$ referring to real and imaginary parts and 
the Kronecker delta function implying the    
non-randomness of off-diagonal  matrix element  
($H_{kl}= b_{kl;1} + i b_{kl;2}$).  
However $\rho(H)$ can be expressed in a more general form: 
\begin{eqnarray}
 \rho (H,y,b)=C{\rm exp}({- \sum_{s=1}^2 \sum_{k\le l} \alpha_{kl;s}
 (H_{kl;s}- b_{kl;s})^2 })   
\end{eqnarray}
 with $C=\prod_{k\le l} \prod_{s=1}^2 \sqrt{\alpha_{kl;s}\over \pi}$ 
as the normalization constant,   
 $y$  as the set  
 of the coefficients  
$y_{kl;s}=\alpha_{kl;s} g_{kl}={g_{kl}\over 2 <H^2_{kl;s}>}$ 
 and $b$ as the set of all $b_{kl;s}$ with $g_{kl}=1+\delta_{kl}$. 
As obvious, 
 in limit 
$\alpha_{kl;1}, \alpha_{kl;2} \rightarrow \infty$, 
eq.(1) corresponds to the  
non-random nature of the off-diagonal part of $H$. 
However note, for later reference, that a same mean value for all the 
matrix elements, 
that is $b_{kl;s}=\epsilon$ for all $k,l$ (irrespective of $\alpha_{kl}$ 
values) implies just a shift of the origin 
in the matrix space and does not affect the eigenvalue distribution.

The MI transition is brought about by a competitive variation of the 
disorder and hopping rate which can be mimicked by a change of the  
distribution parameters of $\rho(H)$. 
The changed  
 eigenvalue distribution as a result can be obtained by integrating 
$\rho(H)$ over associated eigenvector space. 
Let $P(\mu,y,b)$ be the probability of finding eigenvalues 
$\lambda_i $  of $H$ between $\mu_i$ and $\mu_i+{\rm d}\mu_i$ at 
a given $y$ and $b$, it can be expressed as   
$P(\mu,y,b)= \int
\prod_{i=1}^{N}\delta(\mu_i-\lambda_i) \rho (H,y,b){\rm d}H $. 
As discussed in ref.\cite{ps}, a particular combination of the evolution  
of $P$ with respect to various distribution parameters, namely,  
$ {\it D} P \equiv 2\sum_s \sum_{k \le l} y_{kl;s}(\gamma - y_{kl;s})
{\partial P\over\partial y_{kl;s}} -   \gamma
\sum_s \sum_{k \le l} b_{kl;s}{\partial P\over\partial b_{kl;s}}$, 
with $\gamma$ as an arbitrary parameter,     
results in the diffusion of eigenvalues with a finite drift due to their 
mutual repulsion.
\begin{eqnarray}
{\it D} P  &=& 
\sum_n {\partial \over \partial\mu_n}
\left[{\partial \over \partial\mu_n} +
 \sum_{m\not=n}{\beta \over {\mu_m-\mu_n}} + \gamma \mu_n \right] P 
\end{eqnarray}
with $\beta$ depending on the symmetry class of the ensemble 
($\beta=1$ for real symmetric matrices and 
$\beta=2$ for complex Hermitian ones).
 The multi-parametric evolution of $P$ can further be reduced in terms of the  
 evolution with respect to a single parameter $Y$, that is,  
 ${\it D}P = {\partial P\over \partial Y}$ where 
$Y$ is a function of various $y_{kl;s}$ and $b_{kl;s}$ \cite{ps},  

\begin{eqnarray}
Y= {1\over 2 N^2} \sum_{k\le l} \sum_{s=1}^2 \left[ {1\over 2}
 {\rm ln}{y_{kl;s}\over |y_{kl;s}-\gamma|} -
{1\over \gamma} {\rm ln} |b_{kl;s}|\right] +C 
\end{eqnarray}
  
and can therefore be 
referred as the 'complexity parameter'. 
Here $C$ is an arbitrary constant. The eq.(2) can now be rewritten as follows,

\begin{eqnarray}
{\partial P\over\partial Y} 
   &=& 
\sum_n {\partial \over \partial\mu_n}
\left[ {\partial \over \partial\mu_n} +
 \sum_{m\not=n}{\beta \over {\mu_m-\mu_n}} + \gamma \mu_n \right] P 
\end{eqnarray}

For $y_{kl;s} \rightarrow \gamma$ and $b_{kl;s} \rightarrow \epsilon$ 
(for almost all $k,l$), 
$ {\it D} P \rightarrow 0$ and the evolution reaches the 
steady state given by a WD distribution  
$P(\mu) = \prod_{i< j}|\mu_i - \mu_j|^{\beta} 
{\rm e}^{-{\gamma\over 2}\sum_k \mu_k^2}$. (Note, in this limit,  
$\rho(H) \propto {\rm e}^{-\gamma {\rm Tr}H^2 - \epsilon {\rm Tr}H}$ which 
indeed  has a WD distribution and therefore agrees with the solution of 
eq.(4) in limit ${\it D} P \rightarrow 0$). 
The eq.(4) is similar to the equation governing a perturbation 
driven transition of the eigenvalues, of a Hamiltonian 
$H=H_0 +\tau V$, between various universality classes of WD ensembles.  
 Here $H_0$ is an initial ensemble  
 with perturbation $V$ causing a transition to 
another (or same) universality class.  
 The transition to 
equilibrium, with  perturbation strength $\tau$ as the evolution 
parameter, is rapid, 
discontinuous 
for infinite dimensions of matrices (analogous to Anderson 
transition). But for small-$\tau$ 
and large $N$, a smooth transition can be seen in terms of a rescaled 
parameter $\Lambda$ which measures locally the mean-square perturbation  
matrix element in units of the average level spacing \cite{dy}. 
The intermediate states of this transition have been well-studied in past and 
many results for their spectral fluctuations are already known \cite{ap,fkpt}.
 By replacing the perturbation parameter by complexity parameter, 
these results  can directly be used for the corresponding measures for 
the MI transition.   
Note this is eqivalent to say that if the initial Anderson Hamiltonian 
is described by 
$H_0$, the changing complexity (the combined effect of changing disorder 
and hopping rate) 
can be treated as a perturbation 
$V$ of strength $Y$, taken from WD ensemble,  
 with Hamiltonian $H$ of the system
 represented by $H=H_0 + Y V$.

	The $n^{\rm th}$ order level density correlation 
$R_n$ is defined as  $R_n = <\nu (\mu_1,Y)..\nu(\mu_n,Y)>$
with $\nu(\mu,Y) = N^{-1} \sum_i \delta (\mu-\mu_i)$ as the density of 
eigenvalues  and $<..>$ implying the ensemble average. The $R_n$ can 
also be expressed in terms of $P$: 
$R_n = { N! \over {(N-n)!}}\int P(\mu, Y) 
{\rm d}\mu_{n+1}..{\rm d}\mu_N$. A solution of  eq.(4), averaged over 
initial conditions,  can therefore be used 
to obtain various correlations for a given set of distribution 
parameters. In practice, however, the involved integrals could 
only be solved for the cases when $H$ is complex Hermitian in nature 
\cite{ap,ps}.  
 a direct integration of F-P equation (4)  
leads to the BBGKY hierarchic relations among the unfolded correlators 
$R_n(r_1,..,r_n;\Lambda)={\rm Lim} N\rightarrow \infty ;
{{\it R}_n(\mu_1,..,\mu_n;Y) \over 
{\it R}_1(\mu_1;Y)...{\it R}_1(\mu_n;Y)}$
with $r=\int^{r} {\it R}_1(\mu;Y){\rm d}Y$ and 
$\Lambda=(Y-Y_0)\gamma/\Delta^2$ 
($\Delta$ as the mean level spacing $=R_1^{-1}= (N<\nu>)^{-1}$) \cite{fkpt,ap},
\begin{eqnarray}
{\partial R_n \over\partial \Lambda} &=& 
\sum_j {\partial^2 R_n\over \partial r_j^2}
-\beta \sum_{j\not=k} {\partial \over \partial r_j} 
\left({R_n \over {r_j-r_k}}\right) 
-\beta \sum_j {\partial \over \partial r_j} 
\int_{-\infty}^{\infty} {R_{n+1} \over {r_j-r_k}}
\end{eqnarray}
Here note a perturbation of strength $\Lambda$ mixes levels in a 
spectrum span of $\sqrt\Lambda$ only, due to level repulsion \cite{fkpt}. 
This results in a sensitivity of the correlations within a given 
energy range $r$  to the ratio $r/\sqrt\Lambda$. 
In the case of MI transition, therefore, $\Lambda$ can be identified 
with $g^2$.   
 
For $n=2$ and small values of $r$, the integral term in eq.(5) 
makes a negligible contribution thus leading to following approximated 
closed form equation for $R_2$

\begin{eqnarray}
{\partial R_2 \over\partial \Lambda} &=& 
2 {\partial^2 R_2\over \partial r^2}
- 2\beta {\partial \over \partial r} 
{R_2 \over r}
\end{eqnarray}
%
%
Similarly for large-$r$ behaviour, $R_2$ can be obtained from the following 
relation
\begin{eqnarray}
R_2(r,\Lambda)= R_2(r,\infty) + 2\beta \Lambda 
\int_{-\infty}^{\infty} {\rm d}s {R_2(r-s;0)-R_2(r-s;\infty) \over 
s^2 + 4\pi^2 \beta^2 \Lambda^2 }.   
\end{eqnarray}
The hierarchic equation can then be used to obtain an approximate form 
of the higher order 
 correlations. For example, the approximate information about $R_3$ can 
be extracted by a substitution of large and small $r$ behaviour of $R_2$ 
in eq.(5) with $n=2$. 


	For studies of the level-correlations during  MI transition,  
the system can initially be considered in an insulator regime where all the 
eigenvectors become localized on individual sites of the lattice 
(strong disorder limit). This results in a diagonal form of the matrix $H$ 
  with the eigenvalues  
independent from each other. The insulator limit can therefore be modeled 
by ensemble (1) with $\alpha_{kl} \rightarrow \infty$ for $k\not= l$, 
 $\alpha_{kk}=\alpha_I$ (for all $k$-values) 
 and $b_{kl}\rightarrow \epsilon$ (for all $k,l$), giving, 
$Y_0 = {1\over 4 N}{\rm ln} 
{2\alpha_0\over |2\alpha_0-\gamma|} - {1\over  
\gamma} {\rm ln}\epsilon + C$. (Here the origin of the energy-axis is chosen 
at $\epsilon$ so as to avoid the  undefined nature of ${\rm ln}b_{kl}$ near 
$b_{kl} \rightarrow 0$).  
 The decrease of the diagonal disorder, that is,  
 an increase of $\alpha_{kk}$ from $\alpha_0$ to some finite values
(while $\alpha_{kl}$, $k\not= l$, remains infinite throughout the transition)
 will ultimately lead to metal regime with 
fully delocalized wavefunctions. A change in the hopping rate of the connected 
sites from $\epsilon$ to $\epsilon+t$ ($b_{kl} \rightarrow \epsilon +t$ for only those $k,l$ values which are connected, $b_{kl} =\epsilon$ for others) 
will also affect the transition. As mentioned above, 
the eigenvalue distribution of $H$ in the  
metal regime can be well-modeled by the WD ensemble; let it be 
described by  all $\alpha_{kl}\rightarrow \alpha_M$($ > \alpha_0$).
 Thus for the study of transition 
in this case 
we can choose $\gamma=2\alpha_M$.   
For any  
intermediate state of the transition with hopping rate $t$ and disorder 
coefficient $\alpha$, the complexity parameter  
       $Y    = {1\over 2 N^2}\left[ {N\over 2}{\rm ln} 
{2\alpha\over |2\alpha-\gamma|} - 
{ K \over \gamma} {\rm ln} (t+\epsilon) - 
{2 N^2-K\over \gamma}{\rm ln}\epsilon \right] + C$. 
 Here $K = \kappa N$ is the total number of the sites 
connected by hopping and 
depends on the 
dimensionality $d$ of the system (as $N=L^d$).
The transition parameter can now be given as follows, 
with the mean level spacing $\Delta = (<\nu> N)^{-1}$:  
$\Lambda = {2\alpha_M (Y - Y_0)/ \Delta^2} = 
\alpha_M <\nu>^2 N a(\alpha,t)/2$
with $a(\alpha,t) \equiv \left[{\rm ln} 
{\alpha |\alpha_0 -\alpha_M|\over \alpha_0|\alpha-\alpha_M|} - 
{\kappa \over \alpha_M} {\rm ln}\; {t + \epsilon\over \epsilon}
 \right]$.
As obvious from the above, the transition is governed  
by relative values of the disorder and the hopping. Here 
$\Lambda \rightarrow 0$  leads to a fully localized regime which, 
corresponds to  the condition
$a(\alpha,t) 
 < {1\over N}.$
Similarly the condition   
$a(\alpha, t) > {1\over N}$
corresponds to $\Lambda \rightarrow \infty$ for large $N$ 
and therefore  extended states. For $a(\alpha,t) = {\mu \over N}$ ($\mu$ as a 
$N$-independent function), $\Lambda=\alpha_M <\nu>^2 \mu \equiv \Lambda^*$ is independent of 
$N$ which therefore remains 
same for infinite systems ($<\nu>$ being independent of $N$). This   
 $\Lambda$-value corresponds to the critical point of the transition 
in infinite systems, the statistics being Poisson and WD below and above 
$\Lambda^*$, respectively. The  condition for the 
critical region or mobility edge can therefore be given as  
${\rm ln} {\alpha \over \alpha_0} + {\alpha -\alpha_0 \over \alpha_M}  
\approx   {\kappa \over \alpha_M}{\rm ln}\;(t+1) $ 
(taking $\epsilon=1$ without any loss of generality). 
As ${|\alpha -\alpha_0|\over \alpha_M} << 1$ even for large $\alpha$-values,  
 the condition is always satisfied if  
${\kappa\over \alpha_M} \rightarrow 0$. 
This explains the localization of all the  
states in infinitely long wires (or strictly 
1-d systems where $z$ is very small) even for very weak 
disorder. With increasing dimensionality $d$, connectivity $K$ 
of the lattice 
 and thereby the possibility of $\Lambda >> 0$ and the delocalized
 states increases. 
As reflected by its form,  $\Lambda$ also plays the role of  
the finite size scaling parameter. 
Note that the $\Lambda$-dependence of various 
correlators indicates an interaction of levels different from 
that of Poisson or WD if $\Lambda$ remains finite in limit 
$N\rightarrow \infty$.  
 This implies the presence of a new type of universal statistics near the 
critical point $\Lambda^*$  of the transition. 
For finite systems, however, there is a continuous family of 
 intermediate 
 statistics between Poisson and WD, described by $\Lambda$     
and the symmetry of the Hamiltonian.


	The fluctuation measures for various stages of the MI transition 
can be obtained from those of the perturbation driven Poisson 
$\rightarrow $ WD ensemble transition, by replacing perturbation 
parameter by $Y$. 
The two level density correlator $R_2(r;\Lambda)$ for the 
Anderson transition in finite systems, 
in presence of 
a magnetic field, can therefore be given by the $R_2$ for 
Poisson $\rightarrow$ GUE transition driven by the perturbation\cite{ap,ks}: 
\begin{eqnarray}
 R_2 (r;\Lambda) - R_2(r;\infty)={4\over \pi}\int_0^\infty {\rm d}x 
\int_{-1}^1 {\rm d}z \;{\rm cos}(2\pi rx) 
\;{\rm exp}\left[-8\pi^2\Lambda x(1+x+2z\sqrt x)\right]
\left({\sqrt{(1-z^2)}(1+2z \sqrt x) \over 1+x+2z \sqrt x}\right)
\end{eqnarray}
where $R_2(r,\infty)= 1- {{\rm sin}^2(\pi r) \over \pi^2 r^2}$ 
 and $R_2(r,0)=1$ corresponding to metal and insulator regime 
respectively. A substitution of $\Lambda=\Lambda^*$ in eq.(8) 
will thus give the 
two level correlation for critical regime of Anderson transition. 

An important characterstic 
of  CSS is the level compressibility $\chi$ which is basically 
a measure of the level repulsion; $\chi=0$ in the metallic phase and 
$\chi=1$ in the insulator phase and takes an intermediate value between 
$0$ and $1$ at the hybrid phase near the critical point.      
The eq.(4.15) of \cite{ks} can  be used to obtain  
 $\chi(\Lambda) \approx 
1 - \int_{-\infty}^{\infty} Y_2 (r;\Lambda) {\rm d}r$ with $Y_2 =1-R_2$, 
$\chi(\Lambda)= 1 + \int_0^\infty {\rm d}u 
{\Lambda \over z}[I_1(z)-\sqrt{8z/\pi}I_2(z)] 
{\rm exp}[-(\Lambda/ 2) u^2 - \pi\sqrt{\Lambda/2} u] $
with $z=\sqrt{2*\pi \Lambda^2 u^3}$ and $I_n$ as the $n^{\rm th}$ 
Bessel function.
(Note eq.(8) can not directly be used to calculate $\chi$ due to 
technical difficulties).  
It can be checked that $\chi \rightarrow 1$ 
for $\Lambda \rightarrow 0$  (insulator limit) and $\chi \rightarrow 0$ 
for $\Lambda \rightarrow \infty$ (metal limit). As obvious from the smooth 
nature of the functional form of $\chi$, any non-zero finite $\Lambda$-value
will correspond to a value of $\chi$ intermediate between zero 
and one. The critical region will thus have  
a finite level compressibility different from 
both metal and insulator regimes and therefore multifractal nature 
of the eigenvectors \cite{ckl}. 

 The statistical measures for the Anderson transition in 
presence of a time-reversal symmetry can similarly be obtained by using their 
 equivalence  to those for 
the Poisson $\rightarrow$ GOE transition. Although much information is 
not available about the latter (due to the technical 
difficulties in integrating over associated orthogonal space of 
eigenvectors) 
however some approximate results are known.  
For example, the $R_2$ for small $r$  
can be obtained by solving the eq.(6) for $\beta=1$, 
$R_2(r,\Lambda) \approx (2r-1)^{1/2} 
J_{1/3}\left((2r-1)^{3/2}/3\Lambda \right) {\rm e}^{r/2\Lambda}$ 
with $J(z)$ as the Bessel function. For $r<<\sqrt\Lambda$, $R_2 \propto r$ 
implying level-repulsion of the same degree as for the delocalized regime; 
this is consistent with description of $\Lambda$ as $g^2$.  
For large-$r$, $R_2$ can be obtained from eq.(7) by taking  
$R_2(r;0)=1$ (Poisson limit), $\beta=1$ and $R_2(r,\infty)= 
 1- {{\rm sin}^2(\pi r) \over \pi^2 r^2} - 
\left(\int_r^\infty {\rm d}x {{\rm sin}\pi x \over \pi x}\right) 
\left({{\rm d}\over {\rm d}r}{{\rm sin}\pi r \over \pi r}\right)$ 
(GOE limit); $R_2(r,\Lambda) \approx R_2(r,\infty) + 
{4\Lambda \over r^2 + 4\pi^2 \beta^2 \Lambda^2}$ . 

The exact formulation for the 
level compressibility $\chi$ for this case can not be obtained due 
to lack of the knowledge of $R_2(r,\Lambda)$ for entire energy-range. 
However the presence of a non-zero, non-unity compressibility can also be 
seen from the nearest-neighbour spacing distribution $P(s)$ for large $s$ 
values. This is because the compressibility of the spectrum is basically  
a reflection of the correlations of levels at large distances.   
For WD case, $P(s) \propto s {\rm e}^{-\pi s^2/4}$ decays as a 
Gaussian near the tail indicating 
no compressibility of levels. As for Poisson limit 
$P(s)\propto {\rm e}^{-s}$, the unity 
coefficient of $s$ in the exponent implies  a 
very high degree of compressibility. An exponential 
tail of $P(s)$ with a non-unity  coefficient will thus correspond to a 
a compressibility less than Poisson case;    
 if $P(s) \propto {\rm e}^{-ks}$ for large $s$, $\chi \approx 1/2k$.   
The $P(s)$ for MI transition can be given by using the one for the 
Poisson $\rightarrow$ 
GOE transition \cite{lh},
$P(s,\Lambda)= {s u^2(\Lambda)\over \Lambda}
{\rm e}^{-u^2(\Lambda) s^2/4\Lambda^2} 
\int_0^{\infty} {\rm d}\zeta {\rm e}^{-\zeta^2 -2\zeta\Lambda }
I_0(s\zeta u(\Lambda)/\Lambda) $
with $u(\Lambda)=\sqrt\pi U(-1/2,0,\Lambda^2)$, $U$ being Kummer function 
and $I_0$ Bessel function. 
Although this result is rigorous for $2\times 2$ 
matrix space but is proved reliable for systems with many levels. 
(The dominance of nearest-neighbour interaction is also supported by the 
mapping of the eigenvalues dynamics 
 to the particle dynamics of calogero 
Hamiltonian with inverse-square interaction).  
As can 
be checked,for $s >> \Lambda$, $P(s,\Lambda) \propto {\rm e}^{-k s}$ with 
$k=\sqrt{\pi \Lambda}$. Near the critical point, 
therefore, $k= \sqrt{\pi\Lambda^*} > 1$ and 
$\chi \approx 1/2\sqrt{\pi\Lambda^*}$. For 
small separations, that is, $s << \Delta$, $P(s,\Lambda^*)$ has a 
linear dependence on $s$, indicating a WD type behaviour. The 
spacing distribution for CSS is thus a hybrid of Poisson and WD 
type distributions; this result is in agreement with numerical 
study of \cite{ss}.       

The case of the MI transition 
due to off-diagonal Gaussian disorder can similarly  be studied by 
taking finite $\alpha_{kl}$ values for all $k,l$; the $Y$ for any 
intermediate stage of the transition brought about by a most general 
type of diagonal and off-diagonal disorder is given by eq.(3). As obvious 
from the above,  
the presence or absence of the off-diagonal disorder will only affect  
 the form of $Y$, leaving, therefore, the qualitative behaviour of 
various correlators unaffected. The presence of different boundary conditions 
(e.g. periodic) or different topology for disordered lattice will also 
result in different $Y$-values (as various $y_{kl}$s will be affected), leading to different level-statistics 
even if degree of disorder, hopping rate and dimensionality is same; this is 
in agreement with numerical observations \cite{bran}.

          It should be noticed here that the  
 expressions of $\chi(\Lambda)$ as well as $P(s,\Lambda)$ are  
also valid for Rosenzweig-Porter (RP)  
ensemble ($<H_{ii}^2>= v_1$ for all $i$,  $<H_{ij}^2>=v_2$ for 
all $i\not=j$ and $v_1, v_2$ different in general), thus implying a 
non-zero compressibility for finite $\Lambda$-values in RP model too 
contrary to earlier claims \cite{fgm}. Our result is in 
agreement with Dyson's conjecture for Brownian ensembles \cite{dy} 
according to which the parameter $\Lambda$ governs a smooth transition 
in fluctuation properties; the large-$s$ behaviour too, of $P(s,\Lambda)$,    
is therefore expected to vary smoothly from ${\rm e}^{-s}$ (Poisson) to 
${\rm e}^{-\beta s^2}$ (Wigner).

	To summarize, our analytical study shows  
 that the statistical  
 transition of levels of a disordered system due to change of 
the parameters e.g. disorder,
 and system size can be reduced to the transition with respect 
 to a single scaling parameter.    Our 
results also confirm the presence of just three type of universal statitsics
for inifinite system, namely Poisson above critical value of the 
disorder, WD  below critical value of disorder and a new type  
of statistics for the critical region. The critical region statistics  
is a hybrid of the Poisson and WD and shows a finite level compressibility.  
The universality 
of CSS is restricted in the sense that, although system-independent, 
 it depends on the critical 
value $\Lambda=\Lambda^*$  too besides the underlying symmetry 
of the system.  
We have discussed here only the $2^{nd}$ order level-statistics for Anderson 
transition. The information about the higher order correlations can 
be obtained from the hierarchical equation (5) for the correlators 
or by analyzing particle correlators of CSM Hamiltonian. Further the 
mapping of the level-dynamics during Anderson transition to particle-dynamics 
of CSM Hamiltonian once more indicates the dominance of two body interaction 
in nature.



\begin{references}
\bibitem{and}
P.W.Anderson, Phys. Rev. 19, 1492, (1958);
P.A.Lee and T.V.Ramakrishnan, Rev. Mod. Phys. 57, 287 (1985).
\bibitem{ss}
B.I.Shklovskii, B.Shapiro, B.R.Sears, P.Lambrianides and H.B.Shore, 
Phys. Rev. B, 47, 11487 (1993).
\bibitem{fh}
F.Haake, {\bf Quantum Signature of Chaos}, Springer, Berlin (1991).
\bibitem{ckl}
B.L.Altshuler, I.Kh. Zharekeshev, S.A.Kotochigova and B.Shklovskii, 
JETP, 67, 625 (1988); 
J.T.Chalker, V.E.Kravtsov and I.V.Lerner, Pisma Zh. Eksp. Teor. Fiz. 
64, 355 (1996).
\bibitem{klaa}
V.E.Kravtsov, I.V.Lerner, B.L.Altshuler and A.G.Aronov, 
Phys. Rev. Lett. 72, 888 (1994);
V.E. Kravtsov and I.V.Lerner, 
J. Phys.A 28, 3623 (1995).
\bibitem{ps}
P.Shukla, Phys. Rev. E 62, 2098, 2000;  
P.Shukla, to appear in Physica E, 2000 and Physica A, 2000.
\bibitem{dy}
F.Dyson, J. Math. Phys. 3, 1191 (1962).
\bibitem{ap}
A.Pandey, Chaos, Solitons and Fractals, 5, (1995).
\bibitem{fkpt}
J.B.French, V.K.B.Kota, A.Pandey and S.Tomsovic, 
Ann. Phys. (N.Y.), 181, (1988).
\bibitem{ks}
H.Kunz and B.Shapiro, Phys. Rev. E, 58, 400, 1998.
\bibitem{fgm}
K.M.Frahm, T.Guhr, A.Muller-Groeling,  Ann. Phys. (N.Y.) 270, 292 (1998).  
\bibitem{lh}
G.Lenz and F.Haake, 67, 1 (1991).
\bibitem{bran}
D.Braun, G.Montambaux and M.Pascaud, Phys. Rev. Lett. 81, 1062, 1998.
\bibitem{}
* E-Mail : Shukla@phy.iitkgp.ernet.in
\end{references}
\end{document}